\documentclass[letterpaper, 10 pt, conference]{ieeeconf}  

\IEEEoverridecommandlockouts                              
\usepackage{graphicx}
\graphicspath{{figures/}}

\usepackage[utf8]{inputenc}
\usepackage[english]{babel}

\usepackage{amsmath}
\usepackage{amsfonts}
\usepackage{amssymb}
\usepackage{array}
\usepackage{mathtools}
\usepackage{adjustbox}

\usepackage{tikz} 
\usepackage{tkz-berge}
\usepackage{pgfplots} 
\pgfplotsset{compat=newest} 
\usetikzlibrary{arrows,%
                petri,%
                topaths,
                matrix}%
\usetikzlibrary{shapes,arrows,positioning,calc}
\usetikzlibrary{shadows}

\pgfplotsset{compat=1.5}

\tikzset{
block/.style = {draw, fill=white, rectangle, minimum height=3em, minimum width=3em},
tmp/.style  = {coordinate}, 
sum/.style= {draw, fill=white, rectangle, node distance=1cm},
input/.style = {coordinate},
output/.style= {coordinate},
pinstyle/.style = {pin edge={to-,thin,black}
}
}

\DeclareMathOperator*{\argmin}{arg\!\min}
\DeclareMathOperator*{\argmax}{arg\!\max}

\definecolor{pluplugreen}{rgb}{0.39,0.667,0.333}
\definecolor{pluplured}{rgb}{0.863,0.133,0.196}
\definecolor{pluplublue}{rgb}{0,0.62,0.86}

\title{\LARGE \bf
Characterization of Biologically Relevant Network Structures form Time-series Data
}

\author{Zoltan A. Tuza and Guy-Bart Stan
\thanks{}
\thanks{Z. Tuza and G. Stan are with the Bioengineering Department at Imperial College London, South Kensington Campus, London SW7 2AZ, United Kingdom 
        {\tt\small z.tuza@imperial.ac.uk, g.stan@imperial.ac.uk}}%
} 

\begin{document}

\maketitle
\thispagestyle{empty}
\pagestyle{empty}

\begin{abstract}
High-throughput data acquisition in synthetic biology leads to an abundance of data that need to be processed and aggregated into useful biological models. Building dynamical models based on this wealth of data is of paramount importance to understand and optimize designs of synthetic biology constructs. However, building models manually for each data set is inconvenient and might become infeasible for highly complex synthetic systems. In this paper, we present state-of-the-art system identification techniques and combine them with chemical reaction network theory (CRNT) to generate dynamic models automatically. On the system identification side, Sparse Bayesian Learning offers methods to learn from data the sparsest set of dictionary functions necessary to capture the dynamics of the system into ODE models; on the CRNT side, building on such sparse ODE models, all possible network structures within a given parameter uncertainty region can be computed. Additionally, the system identification process can be complemented with constraints on the parameters to, for example, enforce stability or non-negativity---thus offering relevant physical constraints over the possible network structures. In this way, the wealth of data can be translated into biologically relevant network structures, which then steers the data acquisition, thereby providing a vital step for closed-loop system identification. 
\end{abstract}
\section{INTRODUCTION}
One of the goals in the Systems and Synthetic Biology is to characterize possible network structures that can explain the observed data. This is usually done by incorporating the information content of noisy experimental data into parametrized process models \cite{ljung_sysid}. Then, these network structures give us a blueprint for the possible interactions between chemical species. Building on that, one can understand a complex biological process or even manipulated it by other chemical species. Synthetic Biology is particularly successful identifying small interaction networks in nature or even building artificial ones from well understood biological parts such as the repressilator. The need to grow this library of well-characterized biological parts to build more complex interaction networks drove the development of high-throughput data acquisition. This type of data collection soon generates a large volume of data which is incompatible with the current model building methodology. Mainly because of the current practice involves manually curating the data and the possible network structure. As the need for fast characterization of biological parts drove development of high-throughput methods, the abundant data from high-throughput methods drives the need for automatic model building methods. 

Automatic model building is standard practice in, for example, machine learning, but usually, the building blocks of such models are general nonlinear functions, e.g. Gaussian kernels. In contrast to that, in Biochemistry we have a well-understood family of nonlinear functions that can capture the underlying chemical interactions. Even first principle models can be built by these nonlinearities. For example, Chemical Reactions Networks (CRNs) are often used to build such models.

In CRN modeling, it is usually assumed that there is a one-to-one correspondence between the dynamical model and the underlying CRN structure. However, this is only true if one builds the differential equations from the CRN structure. The other way may yield multiple structures that exhibit the same dynamics. Since the later direction is used in system identification, it needs to be carefully investigated.   

The existence of multiple network structures for a given dynamics has been investigated extensively in \cite{Johnston2011conj,acs2016}. However, this investigation was done with the assumption of perfect measurements, and the uncertain case was investigated in \cite{acs2018}. As we show in this paper, this allows us to handle noisy time-series data and to compute network structures using the same computational tools developed for the noiseless case. 

The goal of the paper is to develop a framework that characterizes all possible network structures from time-series data. All possible structure means that it can be proven the set of network structures, computed in our framework, is complete and no other other network structure exists for a given dynamical model. Complementing the underlying optimization problem with further constraints on the dynamics or on the structure helps to reduce the number of possible network structures, thus we can not only add \textit{a priori} knowledge to the identification process but measure its impact on the number of possible network structure. This allows us to compare the effect of different \textit{a priori} knowledge. Moreover, we can even characterize the set of assumptions needed for a unique the network structure from time-series data.
\paragraph*{Notations.} $\mathbb{R}_{\geq 0}$ denotes the nonnegative real numbers, $\mathbb{N}_0$ is the set of integers including zero. $[A]_{ij}$ is the entry in the $i$th row and $j$th column of the matrix $A$. Furthermore, $[A]_{i\cdot}$ is the $i$th row of matrix $A$, and diag$[a]$ is matrix which has the elements of $a$ in the diagonal and the rest is zero. Finally, $A\succeq 0$ is a positive semidefinite matrix and $\text{vec}(A) = [[A]_{1\cdot}, [A]_{2\cdot}, \ldots, [A]_{n\cdot}]^\top$ is transpose of the row expansion of matrix $A\in\mathbb{R}^{n\times m}$. 
\section{BACKGROUND}
The concept of CRNs was introduced by Feinberg during his seminal lectures, and since then it has become a widely used modeling framework \cite{Feinberg:79}; first in process engineering, then later in Systems and Synthetic Biology. 

The dynamics of CRNs can be described by the class of kinetic systems which offer certain algebraic proprieties that lead to several powerful results (see, e.g., \cite{Feinberg1987,Craciun2015,Shinar2010}). There has been a constant effort to characterize the systems theoretical properties of kinetic systems (see, \cite{Craciun2009,perez2012chemical,van2013mathematical} for more details). This paper follows an optimization-based approach which translates the dynamical and structural properties of kinetic systems into the constraint set, see \cite{Johnston2011conj,acs2016} for an overview.

This section defines the system class represented by polynomial Ordinary Differential Equations (ODEs) and defines Chemical Reaction Networks as well; then establishes a connection between the two. Building on these definitions, we can introduce the optimization problems to compute certain graph structures.

Let us define the following polynomial differential equation with state vector $x\in \mathbb{R}^n_{\geq 0}$, and 
\begin{eqnarray}
\label{eqn:kinetic_sys}
  \dot{x} = M\varphi(x), \quad x(0) \in \mathbb{R}^n_{\geq 0},
\end{eqnarray}
where the matrix $M\in\mathbb{R}^{n\times m}$ is the coefficient matrix and $\varphi(x): \mathbb{R}^n_{\geq 0} \rightarrow \mathbb{R}^m_{\geq 0}$ is a monomial-type vector mapping defined as $\varphi_j(x)= \prod_{i=1}^{n} x_i^{[B]_{ij}}, j=1,\ldots,m$ and $B\in \mathbb{N}_0^{n\times m}$. Note that the system in \eqref{eqn:kinetic_sys}, with certain sign constraints in matrix $M$, belongs to the class of nonnegative systems, i.e. $\mathbb{R}_{\geq 0}^n$ is forward invariant (see, e.g. Chapter 9 in \cite{Haddad2010}).

Next, we define chemical reaction networks, which can be characterized by three sets:
\begin{itemize}
  \item a set of species: $\mathcal{S}= \{X_i\  |\  i=1,\ldots,n\}$,
  \item a set of complexes: $\mathcal{C} = \{ C_j \ |\  j=1,\ldots,m\}$, where\\
$C_j = \sum \limits_{i=1}^{n} \alpha_{ji} X_i \qquad j=1,\ldots,m$ \text{and}  \\
$\alpha_{ji} \in \mathbb{N}_0 \qquad \qquad ~~j=1,\ldots,m, \   i=1,\ldots,n$,\\
$\alpha_{ji}$ are called the stoichiometric coefficients,
  \item and a set of reactions: $\mathcal{R} \subseteq \{(C_i,C_j)\  |\  C_i,C_j \in \mathcal{C}\}$, each ordered pair $(C_i,C_j)$ has a reaction rate coefficient $k_{ij} \in \mathbb{R}_{\geq 0}$ so that the corresponding reaction $C_i \rightarrow C_j$ takes place if and only if $k_{ij} >0$.
\end{itemize}
In the rest of the paper, we assume mass action kinetics, but the results summarized in this section have been extended to rational kinetics as well, see e.g. \cite{GaborA2016} for more details.

For computation purposes, we can characterize a CRN by two matrices: the complex composition matrix $Y\in \mathbb{N}_0^{n\times m}$ describes the complexes as follows
\[[Y]_{ij}=\alpha_{ji} \qquad  i=1,\ldots,n, \  j=1,\ldots,m,\]
and the set of reactions is encoded by the Kirchhoff matrix $A_\kappa \in \mathbb{R}^{m\times m}$ as
\begin{eqnarray}
\label{eq:kirchhoff_mtx}  
[A_\kappa]_{ij}= \begin{cases} 
\hfil k_{ji} &\text{ if } i \neq j\\
-\sum \limits_{l=1, l\neq i}^{m} k_{il} & \text{ if } i=j.\\
\end{cases}
\end{eqnarray}

The dynamics of a CRN can be written as a nonnegative polynomial differential equation 
\begin{eqnarray}
\label{eqn:crn_dynamics}
  \dot{x} = Y A_\kappa \psi^Y(x), \quad x(0) \in \mathbb{R}^n_{\geq 0},
\end{eqnarray}
where $x$ represents the concentration vector of the species and the monomial vector mapping $\psi^Y(x)$ is defined as
\begin{eqnarray}
  \psi_j^Y(x) = \prod_{i=1}^{n} x_i^{[Y]_{ij}} \quad j=1,\ldots,m.
\end{eqnarray}
At this point, we can make a connection between a nonnegative polynomial ODE and the dynamics of CRNs as follows:
a nonnegative polynomial ODE in \eqref{eqn:kinetic_sys} can be transformed into the form of \eqref{eqn:crn_dynamics}, i.e. 
\begin{eqnarray}
\label{eqn:dyn_eq_full}
M\varphi(x) = Y  A_\kappa \psi^Y(x), 
 \end{eqnarray} 
 if and only if the following condition is fulfilled
\begin{eqnarray}
\label{eqn:kinetic_logical}
\text{if } [M]_{ij}\leq 0,\text{then } [B]_{ij}>0, \\i=1,\ldots,n, j=1,\ldots,m. \nonumber    
\end{eqnarray}  
If the above condition is satisfied, we call \eqref{eqn:kinetic_sys} a kinetic system because it has at least one CRN realization. This condition also ensures that there are no negative cross-effects in the kinetic system, as it is explained in \cite{Erdi1989}. Furthermore, using this condition a so-called canonical realization can be computed from \eqref{eqn:kinetic_sys} (see \cite{Hars1981} for the details). However, it should be stressed that CRNs with different sets of complexes and reactions can generate the same dynamics \cite{Erdi1989,Craciun2008}.

In this paper, we assume that the set of used complexes is known, which defines the matrix $Y$ and consequently the monomial vector mapping $\psi^{Y}(x)$. Note that the left hand side of \eqref{eqn:dyn_eq_full} is a multivariate polynomial function, while the factorization on the right hand side defines a CRN structure. Therefore, the sets of monomials in $\varphi$ and $\psi^Y$ are not necessarily identical. The reason for this is that the monomials of pure product complexes do not appear in the kinetic equations, but $\psi^Y$ contains the monomials of each complex, even if some of them have zero coefficients in the equations. Naturally, $\psi^Y$ must contain all the monomials of $\varphi$. This means that without the loss of generality we can assume that $\varphi=\psi^Y$ and write the matrix $M$ accordingly. Using this assumption, dynamical equivalence can be simply represented as
\begin{eqnarray}
\label{eqn:dyn_eq}
  M = Y  A_\kappa. 
\end{eqnarray}
The equation above shows that even with a fixed set of complexes, several different $A_\kappa$ matrices can lead to the same dynamics. Hence, these different $A_\kappa$ matrices are called dynamically equivalent realizations of a kinetic system. The total number of such realizations and their (structural and dynamical) properties are the main focus of the past and the current research presented in this paper.

It should be noted that a kinetic system---with a fixed set of complexes---is uniquely characterized by the matrix pair ($Y$, $M$), thus we can refer to it by this pair. 
%
%
%
\subsection{Optimization-based computation of realizations}
\label{sec:opt_based_crn}
Optimization problems can be formulated to find realizations of the same kinetic system with different dynamical or structural properties \cite{acs2016}. Jonhston \textit{et al.} also proved that there exists a unique superstructure for each kinetic system, and all possible realizations are contained by the superstructure as subgraphs \cite{Johnston2012a}. This superstructure is also called the dense realization since it contains the most number of reactions for a given kinetic system.

Based on the superstructure property, an algorithm was developed to compute all possible realizations of a given kinetic system \cite{acs2016}. This algorithm effectively excludes different edge patterns from the dense realization, and by construction, it returns all possible realizations (see details of the proof in \cite{acs2016}). Additionally, this algorithm can be massively parallelized (see \cite{TuzaFosbe2016} for more details).

A kinetic system has not only a structurally unique dense realization but a structurally sparse realization as well. However, this sparse realization is not always structurally unique, meaning that multiple sparse realizations may exist and have the same minimum number of reactions (i.e. edges in the reactions graph).

Several graph properties can be translated as constraints in an optimization problem, a non-exhaustive list includes weak reversibility \cite{Szederkenyi2011b}, complex balance \cite{Szederkenyi2011a}, deficiency zero \cite{Liptak2015}, or deficiency one \cite{Johnston2016}. The resulting optimization problems not only give solutions with the given graph properties, but these may also give a certificate about the lack of such realizations. For example, if one is looking for a weakly reversible realization of a given kinetic system, and there is provably no feasible solution to the optimization problem complemented with the constraints of weak reversibility, then there exists no weakly reversible realization for a given dynamics. Thus, such optimization problems can be used to characterize some of the structural and dynamical properties of kinetic systems.  

\subsubsection{Computation of the dense realization}
The computation of the dense reaction graph can be formulated as an optimization problem. A possible approach for that would be a mixed integer linear programming problem where the number of reactions in the network has to be maximized, see \cite{Szederkenyi2011} for the details. To make the computation tractable for large networks, an iterative algorithm to compute the dense realization was reported in \cite{Acs2015}. The main steps of this algorithm are summarized below.

First, by combining \eqref{eq:kirchhoff_mtx} and \eqref{eqn:dyn_eq}, a linear programming (LP) problem can be formulated with the following set of constrains 
\begin{align}
\label{eq:kinetic_constraints}
\ &  M = Y A_\kappa\nonumber \\
\ &[A_\kappa]_{ij} \geq 0  \quad   i=1,\ldots,m,~~j=1,\ldots,m \  ~~i\neq j\\
\ & [A_\kappa]_{ii} =- \sum \limits_{\substack{j =1 \\ j \neq i}}^{m} [A_\kappa]_{ji} \quad   i=1,\ldots, m,\nonumber
\end{align}
where dynamics of a kinetic system is given by ($Y$, $M$) and $A_\kappa\in\mathbb{R}^{m\times m}$ is now the decision variable of the optimization problem.   

In many cases, the edge exclusion from the reaction graph is needed. Formally, a set $\mathcal{H}\subset \mathcal{R}$ of reactions has to be excluded from the network that can be written as a linear constraint: 
\begin{equation} 
\label{eq:edge_exclusion}
[A_\kappa]_{ji} = 0 \qquad  (C_i,C_j) \in \mathcal{H}.
\end{equation}
Second, we formulate the linear cost function as 
\begin{eqnarray}
\label{eq:cost_fcn}
  \text{maximize}~\sum \limits_{i=1}^m \sum  \limits_{j=1}^m [E]_{ij} [A_\kappa]_{ij},
\end{eqnarray}
where the binary matrix $E\in\{0,1\}^{m\times m}$ selects the elements of $A_\kappa$ into the cost function. Further details of the optimization problem and the algorithm itself is given in \cite{Acs2015}.

\subsection{Uncertain Kinetic Systems}
So far, we encoded dynamics of a kinetic system by ($Y$, $M$) and we assumed that the coefficients matrix $M$ is constant. We then computed certain graph structures. In this section, we define a family of kinetic systems where elements of the coefficient matrix belong to a set, denoted by $\mathcal{M}$ which contains all admissible parameter vectors of the kinetic system. Clearly, the properties of the kinetic system depend on the set $\mathcal{M}$. Therefore, we first characterize the type of uncertainty considered in this paper. Then, building on the results summarized in the previous section, we define a convex optimization problem to compute the dense realization and subsequently all realizations of the uncertain kinetic system. 
\subsection{Optimization-based computation of uncertain realizations} 
\label{sec:uncertain_opt}
We assume two properties of the parametric uncertainty. First, the nominal matrix $\bar{M}\in\mathcal{M}$ is given, thus we have one member of the family of kinetic systems. Second, the nominal $\bar{M}$ is perturbed by an unknown matrix $\Delta$, and we only know the upper bound of the uncertainty in some norm, e.g. in Frobenius norm: $||\Delta||_F \leq \rho$, where $\rho\in\mathbb{R}_{\geq 0}$. With these assumptions, we can characterize the possible parameters vectors as $\text{vec}(M) = \text{vec}(\bar{M}) + \rho u$ where $\text{vec}(M)\in\mathbb{R}^{nm}$ is the vectorization of the parameter matrix $M\in\mathbb{R}^{n\times m}$ and $u\in\mathbb{R}^{nm}$ is given as $u=\text{vec}(\Delta)$. 

The uncertainty set around the nominal $\bar{M}$ is given by  
\begin{eqnarray}
\label{eqn:uncertainty_set_ball}
     \mathcal{M} =\{\text{vec}(\bar{M})+\rho u, ||u||_{2} \leq 1\}. 
\end{eqnarray}   
This type of uncertainty describes a sphere around $\text{vec}(\bar{M})$, which can be translated to an second order conic (SOC) constraint as follows
\begin{eqnarray}
\label{eq:socp_spherical}
	||\text{vec}(M)-\text{vec}(\bar{M})||_2^2 \leq \rho
\end{eqnarray}
where $\text{vec}(M)$ is an optimization variable. We can add \eqref{eq:socp_spherical} as an SOC constraint to the optimization problem defined in \eqref{eq:kinetic_constraints}, \eqref{eq:edge_exclusion}, \eqref{eq:cost_fcn} and compute the dense realization with spherical uncertainty using the Algorithm 1 from \cite{Acs2015}. 

In this paper, we are more interested in the case where the uncertainty not uniform in all directions. In this case, the ellipsoidal uncertainty is defined by $\Sigma\in\mathbb{R}^{nm\times nm}$ and $\Sigma \succeq 0$ and the corresponding uncertainty set is the following 
\begin{eqnarray}
  \mathcal{M} =  \{\text{vec}(\bar{M})+ R u, ||u||_2 \leq 1\},\nonumber
\end{eqnarray}
where $R\in \mathbb{R}^{nm\times nm}$ is defined by the Cholesky decomposition, $\Sigma = R^\top R$. Then, using the same derivation as above, the modified $\mathcal{M}$ can be represented as a SOC constraint 
\begin{eqnarray}
\label{eqn:SOC_with_cov}
   ||R^\top (\text{vec}(M)-\text{vec}(\bar{M}))||_2^2 \leq 1.
\end{eqnarray} 

Again, just as in the spherical case, we can compute the uncertain dense realization. For the complete treatment of uncertain kinetic systems and the proofs, see \cite{acs2018}.
It should be mentioned that other types of uncertainty are possible for a kinetic system, as long as the uncertainty set can be translated as convex constraints, the dense realization exists, see \cite{acs2018}.

Besides the framework described in this paper, a useful application of this technique could be the design of dynamics, i.e. designing a kinetic system which operates inside the prescribed operational limits or design envelope, see \cite{liptak2016kinetic} for details on CRN controller design.

At this point, we have the tools to compute the uncertain dense realization. As it was shown in \cite{acs2018} that the all possible realizations can be computed in the uncertain case as well.

In summary, in order to calculate the dense and subsequently all realizations, we need to define $Y$, the nominal coefficient matrix $\bar{M}$ and a spherical or ellipsoidal uncertainty. Therefore, the next step is to compute $\bar{M}$ from time-series data with the assumption of that $Y$ is known.
%
%
\section{Parameter estimation}
\label{sec:paramest}
There are many possible ways to estimate the parameters of a kinetic system from time-series data, see e.g. \cite{August2009} or \cite{wei2017phd}. In this paper, we work with the following assumptions: all state variables can be measured, and the set of complexes (i.e. the matrix $Y$) is known \textit{a priori}. The former assumption can be relaxed, by using state estimation for the unmeasured states. However, $Y$ is usually assumed to be known, because it represents our knowledge about the participating chemical complexes.

For the purpose of parameter estimation, we need to discretize the kinetic system in \eqref{eqn:kinetic_sys}. Using sufficiently small sampling time, we apply the forward Euler method and get
\begin{eqnarray}
 x_i(t_k) &=& x_i(t_{k-1}) + h[M]_{i,\cdot}\cdot\psi_i^Y(x(t_{k-1})),\\
 x(0)&=&x_0,~~k=1,\ldots,N,~~i=1,\dots,n, \nonumber
\end{eqnarray} 
where $t_k$ is the sampling time point, $x_i(t_k)$ is the $i$th state variable at time $t_k$, $x_0\in\mathbb{R}^n_{\geq 0}$ are the initial values, the $\psi^Y_i(x)$ is the $i$th element of vector mapping $\psi^Y(x)$, the vector $[M]_{i,\cdot}$ is the $i$th row of matrix $M$, the $h$ is the sampling time, and the $N$ is the last sampling time point. 

For the framework later on, we need the time derivate of $x$. It can be estimated in many ways (see \cite{debrabanter2013} for details). In our case, it is given from the forward Euler method. Therefore, we assume that the measurement of $i$th state variable of the discrete kinetic system is available in this transformed form
\begin{eqnarray}
  y^{(i)}(t_k) := \frac{x_i(t_k) - x_i(t_{k-1})}{h},~k=1,\ldots,N,~i=1,\dots,n. \nonumber
\end{eqnarray}
Then, we get a linear process for each state variable which is linear in parameters, and the $i$th state is given as 
\begin{eqnarray}
\label{eq:lin_model}
  \tilde{y}^{(i)}(t_k,\theta) = \Phi(t_{k-1})\theta^{(i)\top} +\nu_i(t_k),
\end{eqnarray}
where the parameter vector is defined as $\theta^{(i)}=[M]_{i,\cdot}$ and the regressor vector is given as
\begin{eqnarray}
  &&\Phi(t_{k-1})=   \\
  &&        \begin{bmatrix}
          \psi_1(x(t_{k-1})), & \psi_2(x(t_{k-1})), & \ldots, & \psi_m(x(t_{k-1}))
          \end{bmatrix}. \nonumber
\end{eqnarray}
and the measurement noise is $\nu_i \sim \mathcal{N}(0,\sigma^2)$. We assume that the distribution of measurement noise is the same for all the output channels. 

It must be emphasized that kinetic system with mass action kinetics is always linear in parameters, therefore the standard algorithms and tools for analysis from the parameter estimation literature can be applied in this case \cite{ljung_sysid}. For example, we can use the well-known Least Squares method to calculate the parameters of \eqref{eq:lin_model} as the following
\begin{eqnarray}
\label{eqn:lsq_with_eq}
  \hat{\theta}^{(i)} &=& \argmin_{\theta} \frac{1}{2}|| y^{(i)} - \Phi\theta ||_2^2 
\end{eqnarray}

The problem with this path is that the Least Squares method does not promote sparsity. In fact, it rather tries to associate non-zero value to all parameters. However, the dynamics of a state variable is usually not driven by all the monomials in $\varphi(x)$, but only a subset of them. 

Therefore, we need to have either a constrained parameter estimation method, which knowns \textit{a priori} the zero coefficients or a parameter estimation method that promotes sparsity. Among many candidates for the later one, Sparse Bayesian Learning gained popularity recently, mostly because of guarantees for convergence and sparsity. However, evolutionary computation \cite{Schmidt2009} or heuristic based \cite{Brunton2016} algorithms were also proposed recently to find parsimonious models from time-series data. 
%
%
%
\subsection{Sparse Bayesian Learning}
\label{sec:SBL}
Sparse Bayesian Learning was proposed by Tipping and was applied to Relevance Vector Machines where the task is to find a sparse regression or classification \cite{Tipping2001}. Independently from the Bayesian framework, Candes \textit{et al.} developed a framework that uses iterative reweighting of the L1 norm penalty on the parameters \cite{Candes2008}. Candes \textit{et al.} makes the connection to MM algorithms, which is a fundamental way to iteratively solve non-convex optimizations problems. Recently, Wipf \textit{et al.}, building on the work of Candes and Tipping developed a framework which uses iterative reweighting of either L1 or L2 norm to find sparse solution of broad range of problems, e.g. sparse signal representation \cite{Wipf2010}, \cite{Wipf2011}, automatic relevance determination \cite{Wipf2008}, source localization on MRI measurements \cite{Wipf2009}. 

Here, we only give a short outline of the Sparse Bayesian framework, and therefore readers are strongly encouraged to read \cite{Wipf2011} and \cite{Wipf2010} for a thorough treatment of the subject. The following introduction follows the notations from \cite{Wipf2010}.

Throughout the derivation, we assume that we have the following process model
\begin{eqnarray}
	y = \Phi\theta + \nu 
\end{eqnarray}
where $y\in\mathbb{R}^{N}$ is the measurement vector, $\Phi\in\mathbb{R}^{N\times m}$ is a dictionary of features, $\theta\in\mathbb{R}^{m}$ is the parameter vector, and $\nu \sim \mathcal{N}(0,\lambda I)$ is the measurement noise. Our goal is to estimate the sparsest $\theta$, which then select the sparest dictionary to describe the measurements.    

As first step in the derivation, the Least Squares problem in \eqref{eqn:lsq_with_eq} can be transformed into a Gaussian likelihood as
\begin{eqnarray}
	p(y|\theta) \propto \exp\left(-\frac{1}{2}||y-\Phi\theta||_2^2\right)
\end{eqnarray}

The prior information about the parameters is expressed in terms of non-negative latent variables $\gamma\in \mathbb{R}_{\geq 0}^{m}$ as follows
\begin{eqnarray}
	p(\theta) \propto \prod_{i=1}^{m} p(\theta_i),\qquad p(\theta_i) = \max_{\gamma_i \geq 0} \mathcal{N}(\theta_i;0,\gamma_i)\zeta(\gamma_i)
\end{eqnarray}
where $\zeta(\gamma_i)$ is a non-negative function. This form of the prior distribution allows us to express wide range of penalty functions that are needed to promote sparsity in the parameter vector \cite{palmer2006variational}. This structure already hints that if we can set the variance of some of the parameters to zero, then the corresponding parameter value becomes zero. The derivation of the algorithm can be found in the Appendix.

To find the parameter values for $\theta$ and $\gamma$, we need to solve the following iterated optimization problem
\begin{itemize}
	\item Step 1: initialize each $z_{i}=1, i=1,\ldots,m$
	\item Step 2: $\hat{\theta} = \argmin_\theta ||y-\Phi \theta||_{2}^{2}+2\lambda \sum_{i}z_i^{-1/2}|\theta_i|$
	\item Step 3: compute $\gamma^{opt}_i = z_i^{-1/2}|\hat{\theta}_i|, i=1,\ldots,m$
	\item Step 4: compute $z^{opt} = \nabla_\gamma \log |\Sigma_y|$
	\item Step 5: iterate Step 2, 3 and 4 until $\gamma$ is converged to some value.   
\end{itemize}
It should be noted that the first iteration of the above algorithm is the LASSO optimization, and we try to improve on that, hence the name iterated reweighted L1 optimization.
%
%
\section{Characterizing network structures}
\begin{figure}
\centering
\begin{tikzpicture}[auto, node distance=3cm,>=latex']
    \node [input, name=rinput] (rinput) {};

    \node [block, below of=rinput, text width=1cm,node distance=1.2cm] (param_est) {\footnotesize Param estim. w/ SBL};
    \node [block, right of=param_est, text width=1.95cm] (dense) {\footnotesize Comp. the dense realization};
    \node [block, below of=dense, text width=1.85cm, , node distance=1.5cm] (core) {\footnotesize Struc. or dyn. constraints};
    \node [block, right of=dense, text width=1.8cm,node distance=3.3cm] (all_real) {\footnotesize Comp. of all realizations};
    \node [output, below of=all_real,node distance=1.2cm] (routput) {};
    \draw [->] (rinput) -- node{$\mathcal{D}$} (param_est);
    \draw [->] (param_est) -- node{$\bar{M},\Sigma$} (dense);
    \draw [->,dashed,pos=0.1] (core)  -|  ++(1.5,1) |-  node[xshift=0.4cm, yshift=-1.3cm]{}  (all_real.200);
    \draw [->,dashed] (core) -| node{} (param_est.south);
    \draw [->,pos=0.41] (dense) -- node{$A_k^D$} (all_real);
     \draw [->] (all_real) -- node[xshift=-1.195cm, yshift=-0.8cm]{$\overbrace{A_k^{(1)},\ldots,A_k^{(P)}}$} (routput);
     \draw [->,dashed,pos=0.7,above] ([xshift=1.2cm, yshift=-0.5cm]  routput) -|  ++(0.15,1) |-  node{further \textit{a priori} information} ([xshift=0.5cm, yshift=-0.32cm] rinput);
    \end{tikzpicture}
\caption{The figure depicts the overall scheme of the framework developed in this paper. $\mathcal{D}$ denotes the time-series data collected from experiments. Then, this information is used to estimate the parameter of the model $\bar{M}$, along with the covariance matrix $\Sigma$. From these, the uncertain dense realization is computed by Algorithm 1 in \cite{Acs2015}. In the final step, all possible realizations are computed and this set can be further analyzed. The framework allows us to complement either the parameter estimation or the structure computation with constraints, thus the possible network structures can be further reduced.}
\label{fig:overall_scheme}
\end{figure}
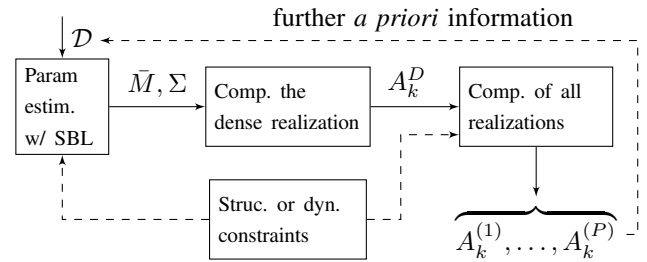

At this point, we can merge the Sparse Bayesian Learning algorithm and the computational tools developed for CRNs into a framework which is shown in Figure \ref{fig:overall_scheme}. As a first step, time-series data are feed to the parameter estimation step (see, Section \ref{sec:SBL}), then using the optimization problems from Section \ref{sec:uncertain_opt}, the uncertain dense realization is computed. From this, all possible graph structures are computed. Thus, this framework transforms the available information in the time-series data into possible network structures relying only on few assumptions. 

Both parts of the framework solve optimization problems. Therefore these optimization problems can be complemented with constraints enforcing structural or dynamical properties.

To visulize our the results, the so-called Feinberg-Horn-Jackson graph is used, which is a weighted directed graph. In this graph, the vertices are the complexes, the edges are the reactions, and the weights are the reaction rate coefficients ($k_{ij}$).
\subsection{Examples}
\begin{figure}
\begin{center}
\begin{tikzpicture}[scale=0.6,transform shape]

 \tikzstyle{VertexStyle}=[shape        = rectangle,
                             fill         = gray!10,
			     draw,drop shadow={opacity=0.2}]
  \Vertex[x=2.8,y=4,L={$X_1$},]{C1}
  \Vertex[x=5.2,y=5.5,L={$2X_2$}]{C2}
  \Vertex[x=5.2,y=1.5,L={$X_1+X_3$}]{C3}
  \Vertex[x=0,y=1.5,L={$X_4$}]{C4}
  \Vertex[x=0,y=5.5,L={$X_2+X_5$}]{C5}
  \node[draw=none,fill=none,opacity=.4] at ($(C1.north) + (0,.6em)$) {C1};
  \node[draw=none,fill=none,opacity=.4] at ($(C2.north) + (0,.6em)$) {C2};
  \node[draw=none,fill=none,opacity=.4] at ($(C3.north) + (0,.6em)$) {C3};
  \node[draw=none,fill=none,opacity=.4] at ($(C4.north) + (0,.6em)$) {C4};
  \node[draw=none,fill=none,opacity=.4] at ($(C5.north) + (0,.6em)$) {C5};
\tikzstyle{LabelStyle}=[fill=none,sloped]
\tikzstyle{EdgeStyle}=[post,sloped,above,inner sep=0pt,-latex]

\Edge[label=$k_{1,2}$,color=blue](C1)(C2)
\Edge[label=$k_{5,2}$,color=blue](C5)(C2)
\Edge[label=$k_{5,1}$,color=blue](C5)(C1)
\Edge[label=$k_{1,3}$,color=blue](C1)(C3)

\Edge[label=$k_{4,3}$,color=blue](C4)(C3)

 \tikzstyle{EdgeStyle}=[post, bend left,sloped,above,inner sep=0pt]
 \Edge[label=$k_{3,5}$,color=blue](C3)(C5)

\end{tikzpicture} 
\begin{tikzpicture}[scale=0.6,transform shape]

 \tikzstyle{VertexStyle}=[shape        = rectangle,
                             fill         = gray!10,
			     draw,drop shadow={opacity=0.2}]
  \Vertex[x=2.8,y=4,L={$X_1$},]{C1}
  \Vertex[x=5.2,y=5.5,L={$2X_2$}]{C2}
  \Vertex[x=5.2,y=1.5,L={$X_1+X_3$}]{C3}
  \Vertex[x=0,y=1.5,L={$X_4$}]{C4}
  \Vertex[x=0,y=5.5,L={$X_2+X_5$}]{C5}
  \node[draw=none,fill=none,opacity=.4] at ($(C1.north) + (0,.6em)$) {C1};
  \node[draw=none,fill=none,opacity=.4] at ($(C2.north) + (0,.6em)$) {C2};
  \node[draw=none,fill=none,opacity=.4] at ($(C3.north) + (0.9em,.6em)$) {C3};
  \node[draw=none,fill=none,opacity=.4] at ($(C4.north) + (0,.6em)$) {C4};
  \node[draw=none,fill=none,opacity=.4] at ($(C5.north) + (0,.6em)$) {C5};
\tikzstyle{LabelStyle}=[fill=none,sloped]
\tikzstyle{EdgeStyle}=[post,sloped,above,inner sep=0pt,-latex]

\Edge[label=$k_{1,2}$,color=blue](C1)(C2)
\Edge[label=$k_{5,2}$,color=blue](C5)(C2)
\Edge[label=$k_{5,1}$,color=blue](C5)(C1)

\Edge[label=$k_{4,3}$,color=blue](C4)(C3)

\Edge[label=$k_{3,2}$,style=dashed](C3)(C2)
\tikzstyle{EdgeStyle}=[post,sloped,above,inner sep=0pt,-latex,pos=0.4]
\Edge[label=$k_{4,1}$,style=dashed](C4)(C1)

\tikzstyle{EdgeStyle}=[post, bend left,sloped,above,inner sep=0pt]
\Edge[label=$k_{1,3}$,color=blue](C1)(C3)
\Edge[label=$k_{3,1}$,style=dashed](C3)(C1)
\tikzstyle{EdgeStyle}=[post, bend left,sloped,above,inner sep=0pt,pos=0.8]
\Edge[label=$k_{3,5}$,color=blue](C3)(C5)
\end{tikzpicture} 
\end{center}
\caption{Left: this example network is taken from \cite{August2009}. The parameters of the network are
$k_{1,2} = 0.3386$, $k_{1,3} = 0.8244$, $k_{5,1} = 0.8496$, $k_{5,2} = 0.4290$, $k_{3,5} = 0.7364$ and $k_{4,3} = 0.5630$. Right: the dense realization computed from data with $\sigma^2=10^{-4}$ and confidence level $\alpha=0.05$. The parameters of the network are $k_{1,2}=0.2920$, $k_{5,2}=0.2645$, $k_{5,1}=0.9208$, $k_{1,3}=1.2262$, $k_{4,3}=0.6108$, $k_{3,5}=0.6495$, $k_{3,1}=0.3227$, $k_{3,2}=0.1800$ and $k_{4,1}=0.5714$.}
\label{fig:acc_network}
\end{figure}
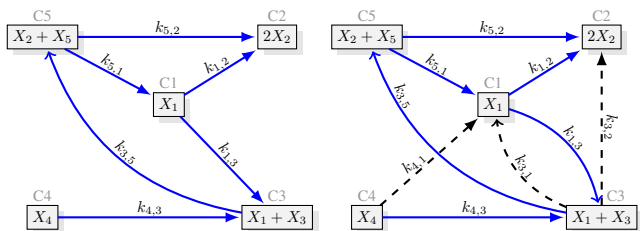

We illustrate the framework shown in Figure \ref{fig:overall_scheme} on an example. This example was presented in \cite{August2009} as a benchmark problem for network inference. The network structure is shown in Figure \ref{fig:acc_network} and its CRN formulation is given as
\begin{small}
\begin{eqnarray}
  Y = \begin{bmatrix}
      1 & 0 & 1 & 0 & 0\\
      0 & 2 & 0 & 0 & 1\\
      0 & 0 & 1 & 0 & 0\\
      0 & 0 & 0 & 1 & 0\\
      0 & 0 & 0 & 0 & 1
  \end{bmatrix} 
\end{eqnarray} 
\end{small}
and
\begin{small}
\begin{eqnarray}
\label{eq:example_ak_param_est}
A_k &=& 
  \begin{bmatrix}
    -1.163  & 0 &  0      &  0       &  0.8492  \\
     0.3386 & 0 &  0      &  0       &  0.4290  \\
     0.8244 & 0 & -0.7364 &  0.5631  &  0       \\
     0      & 0 &  0      & -0.5631  &  0       \\
     0      & 0 &  0.7364 &  0       & -1.2782
  \end{bmatrix},
\end{eqnarray}
\end{small}
the CRN has 5 complexes and 6 reactions, the coefficient matrix is given as $M=Y A_{\kappa}$. The dynamically equivalent (noiseless case) dense realization has also 6 reactions and the network has no other dynamically equivalent realization, i.e. in the noiseless case only one network structure exists. 
\subsubsection{Least Squares Method}
In the first part, we test the performance of the simple Least Squares method assuming that the zero elements of matrix $M$ are known, i.e. we have a constrained Least Squares with equality constraints (LSE). For this part, we generated 50 experiments with different initial values, sampled with Latin Hypercube sampling between $[0,1]$, and simulated the process for $T=10$ sec. For each state variable, we had the following measurement noise $\sigma^2=10^{-4}$, and the sampling time was $h=0.01$. 

From the LSE, the covariance matrix is available in close form. We then set the confidence level at $\alpha=0.05$ and computed the dense realization of the uncertain kinetic system with Algorithm 1 from \cite{Acs2015}; the dense realization is shown in the right panel of Figure \ref{fig:acc_network}. This realization has 9 reactions, thus the extra three reactions open up the possibility for multiple realizations. After computing all possible realizations using the algorithm from \cite{TuzaFosbe2016}, we got 56 structurally different reaction networks.
%
%
%
Clearly, even in this generous scenario---all state variables are directly measured, limited measurement noise is added---gave us several different network structures within the given uncertainty bounds.

\begin{figure}
\begin{tikzpicture}
\begin{axis}[xmode=log, 
        scale only axis=true,
         width=0.4\textwidth,
         height=0.22\textwidth,
          extra y ticks={511},
          extra tick style={grid=major},
          ytick ={10,50,100,200,300,400},
          xlabel=\textsc{$\sigma^2$ - $\log$ scale},
          ylabel=\textsc{Number of realizations}]
\addplot[only marks,scatter,scatter src=explicit] table [x index=1,y index=3, meta index=0, col sep=comma] {noise_vs_realnum_paramest_ACC_model.csv};
\end{axis}
\end{tikzpicture}
\caption{Number of realizations depending on the variance of the measurement noise. LSE case.}
\label{fig:level_vs_real_num}
\end{figure}
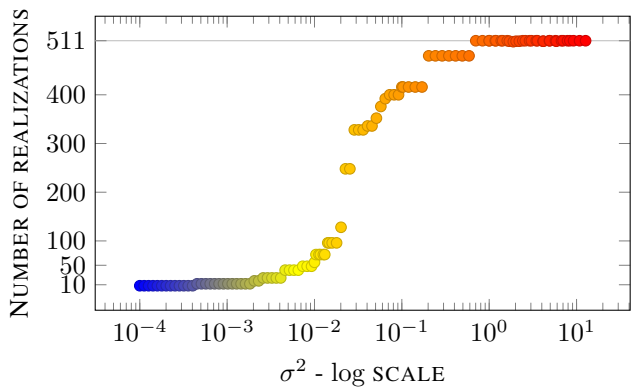

As a next step, we show for this example how the number of possible structures depend on the measurement noise. For that reason, we generated 100 different $\sigma^2$ values between $10^{-4}$ and $10^{1}$. Then, the overall procedure from Figure \ref{fig:overall_scheme} was done for all the different noise scenarios. The Figure \ref{fig:level_vs_real_num} shows the results.
As we can see in Figure \ref{fig:level_vs_real_num}, the number of realizations saturates above a given noise level. This is because we have reached the combinatorially possible number of realizations, which is given by 
\begin{eqnarray}
\label{eqn:comb_max}
   R_{max} = \sum_{i=1}^{R_d} \begin{pmatrix}
     R_d\\ 
     i
   \end{pmatrix}
 \end{eqnarray} 
where $R_d$ is the number of edges in the dense realization. If we have some prior information on the minimum number edges in the network, then the index $i$ can start from that number. Clearly, then we have fewer number of possible network structures.

If we reach the maximum number of possible realizations for a kinetic system that means that the parameters estimation provided no restriction on the possible structures. Therefore, by computing $\frac{R_{\lambda}}{R_{max}}$, where $R_{\lambda}$ is the number of realizations for a given noise level, we have a simple measure which tells us how much information we gained from the parameter estimation about the possible network structures.       

\subsubsection{Sparse Bayesian Learning}
In the second part, we test the performance of the SBL algorithm. The example is the same as before, but this case we did not assume we know the zero entries of $M$ by using equality constraints; it will be estimated by the SBL algorithm. We have generated only $10$ different initial values and simulated the system for $T=10$ sec with sampling time $h=0.1$. Thus, we have significantly less measurement data for all state variables, then in the previous case. 

\begin{figure}
\begin{tikzpicture}
  \begin{axis}[%
        scale only axis=true,
         width=0.4\textwidth,
         height=0.15\textwidth,
  xmode=log,
  legend columns=2, 
  legend style={at={(0.5,-0.2)},anchor=north},
  scatter/classes={%
    a={mark=square*,blue},%
    b={mark=triangle*,red},%
    c={mark=o,draw=blue},
    d={mark=x,draw=black}},
    ytick ={5,7,10,11},
    ylabel absolute, ylabel style={xshift=-0.8cm,yshift=-0.3cm},
     xlabel=\textsc{$\sigma^2$ - $\log$ scale},           
     ylabel=\textsc{Number of realizations}]
  \addplot[scatter,only marks,%
    scatter src=explicit symbolic]%
  table[meta=label] {
x       y      label
0.000100  11   a 
0.000100  5    b 
0.000100  7    c 
0.000100  7    d 
0.000316  11   a 
0.000316  5    b 
0.000316  7    c 
0.000316  7    d 
0.001000  10   a 
0.001000  5    b 
0.001000  7    c 
0.001000  7    d
0.001000  7    d 
0.00316   11   a 
0.00316   5    b 
0.00316   7    c 
0.00316   7    d 
0.01000   10   a 
0.01000   5    b 
0.01000   7    c 
0.01000   7    d
0.0316   11   a 
0.0316   5    b 
0.0316   7    c 
0.0316   7    d 
0.1000   11   a 
0.1000   5    b 
0.1000   7    c 
0.1000   7    d
0.316   11   a 
0.316   5    b 
0.316   7    c 
0.316   7    d 
};
\addlegendentry{$\mathcal{H}=\emptyset$}
\addlegendentry{$\mathcal{H}=[C_4\rightarrow C_1]$}
\addlegendentry{$\mathcal{H}=[C_3\rightarrow C_1]$}
\addlegendentry{$\mathcal{H}=[C_3\rightarrow C_2]$}
  \end{axis}
\end{tikzpicture}
\caption{Number of realizations depending on the variance of the measurement noise in the SBL case.}
\label{fig:level_vs_real_num_SBL}
\end{figure}
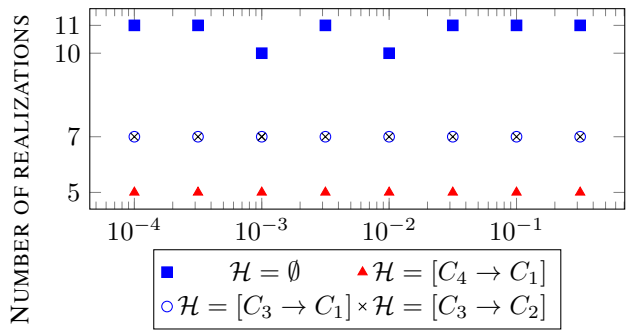

Again, we generated different measurement noise levels and executed the overall procedure from Figure \ref{fig:overall_scheme}. The results of this part is summarized in Figure \ref{fig:level_vs_real_num_SBL}. As we can see, we have significantly fewer possible realizations in the investigated range. In this range of noise level the position of the zero elements in the $M$ matrix was correctly estimated, outside this range the sparsity pattern was not estimated correctly. Additional measurements could potentially extend the range where the sparsity pattern can be restored, but this falls outside of the scope of the current paper.

At this stage, we completely characterized the possible network structures. By adding different \textit{a priori} knowledge, we can measure the effect of this knowledge on the number of structures and compare them with each other. From the dense realization, shown in right panel of Figure \ref{fig:acc_network}, we identify three reactions that are possible reactions given the data, but the are not part of the original network. By excluding these reaction one at the time, we can compare the resulting number of realizations. This is shown in Figure \ref{fig:level_vs_real_num_SBL}, shown with different symbols. 
%
%
%
\section{CONCLUSIONS AND FUTURE WORKS}
We have developed a framework that computes the possible network structures from time-series data. Assuming that all state variables are measured and the participating chemical complexes are known, we have used the Sparse Bayesian Learning to estimate the parameter of the dynamical model. From the statistics of the parameter estimation all possible network structures have been computed. We have given a simple metric to judge the information content of the time-series data as a ratio of the current number of network structures and combinatorially possible ones. We have also shown that we can control the number of network structures by adding additional {a priori} information. In fact, we can compare the effect of different {a priori} information on the number of possible structures, since without any constraints the framework---by construction---provides all possible network structures. As future work, we would like to further characterize data and noise dependence of the Sparse Bayesian Learning algorithm and add further constraints to limit the possible network structures or automatically generate the set of constraints needed for a unique network structure.
\section{ACKNOWLEDGMENTS}
This work was developed within the project called COSY-BIO (Control Engineering of Biological Systems for Reliable Synthetic Biology): H2020-FETOPEN-2016-2017, project number 766840.  This work was also supported by the U.K. EPSRC Fellowship EP/M002187/1 to Dr Stan. The first author is thankful for Gabor Szederkenyi for his advises on the CRN definitions.
\appendix
\section{Appendix}
\label{lab:app}

Derivation of the Sparse Bayesian Learning algorithm using iterative reweigthed L1 minimalization.

For a fixed $\gamma$, we have an approximate prior as
\begin{eqnarray}
	\hat{p}_{\gamma}(\theta) = \prod_{i=1}^{m}\mathcal{N}(\theta_i;0,\gamma_i)\zeta(\gamma_i).
\end{eqnarray}
Using the Gaussian prior and likelihood, we get an approximate posterior
\begin{eqnarray}
	\hat{p}_{\gamma}(\theta|y) = \frac{p(y|\theta)\hat{p}_{\gamma}(\theta)}{\int p(y|\theta)\hat{p}_{\gamma}(\theta) d\theta} =  \mathcal{N}(\theta;\mu_{\theta},\Sigma_{\theta})
\end{eqnarray}
where 
\begin{eqnarray}
\label{eqn:cov_mtx}
	\mu_{\theta} &=& \Gamma\Phi^\top (\lambda I + \Phi\Gamma\Phi^\top)^{-1}y \\
	\Sigma_{\theta} &=& \Gamma - \Gamma\Phi^\top (\lambda I + \Phi\Gamma\Phi^\top)^{-1}\Phi\Gamma
\end{eqnarray}
where $\Gamma = \text{diag}[\gamma]$. The next step is estimating $\gamma$ in a way the is amenable to the above computation.  According to \cite{Wipf2011}, an estimator for $\gamma$ can be constructed using variational representation which involves solving
\begin{eqnarray}
	\gamma &=& \argmin_{\gamma} \int p(y|\theta)|p(\theta) - \hat{p}_{\gamma}(\theta)| d\theta \\
	&=& \argmax_\gamma \int p(y|\theta) \prod_{i=1}^{m}\mathcal{N}(\theta_i;0,\gamma_i)\zeta(\gamma_i) d\theta_i.
\end{eqnarray}
The above expression has analytical solution, after applying $-2\log(\cdot)$, the cost function is 
\begin{eqnarray}
\label{eq:gamma_cost_fcn}
	C(\gamma) = y^\top \Sigma_y^{-1} y + \log|\Sigma_y| + \sum_{i=1}^{m}f(\gamma_i)
\end{eqnarray}
where $f(\gamma_i)=-2\log(\zeta(\gamma_i))$ and $\Sigma_{y}=\lambda I + \Phi\Gamma\Phi^{\top}$.

The above cost function in not convex, but an iterative optimization can be established mainly using results from convex analysis. First, the data dependent term in \eqref{eq:gamma_cost_fcn} can be reexpressed as
\begin{eqnarray}
 	y^\top \Sigma_y^{-1} y = \min_\theta \frac{1}{\lambda}||y-\Phi\theta||_2^2 + \sum_{i=1}^{n} \frac{\theta_i^2}{\gamma_i}.
\end{eqnarray} 
Then, we can create a strict upper bounding function on $C(\gamma)$ by  
\begin{eqnarray*}
	C(\gamma,\theta) = \frac{1}{\lambda}||y-\Phi\theta||_2^2 + \sum_{i=1}^{n} \frac{\theta_i^2}{\gamma_i} + \log |\Sigma_y| + \sum_{i}f(\gamma_i).
\end{eqnarray*}
If we minimize over $\gamma$, then
\begin{eqnarray}
	C(\theta) = \min_{\gamma \geq 0}C(\gamma,\theta) = ||y-\Phi\theta||_2^2 + \lambda g(\theta)
\end{eqnarray}
where 
\begin{eqnarray}
\label{eq:g_fcn}
	g(\theta) = \min_{\gamma \geq 0}  \sum_{i=1}^{n} \frac{\theta_i^2}{\gamma_i} + \log |\Sigma_y| + \sum_{i}f(\gamma_i).
\end{eqnarray}
In the rest of the paper, we assume that $f(\gamma_i)=0$, for $i=1,\ldots,m$ because we are interested in maximal sparsity. For further details on $f(\gamma_i)$ and comparison of different regularization terms, see \cite{wipf2007performance}.

In the rest of the analysis, we focus on \eqref{eq:g_fcn}. The $\log |\cdot|$ is a concave function on semidefinite matrices \cite{Boyd2004} and $\Sigma_y$ is a positive semidefinite matrix and an affine function in $\gamma$. Thus $\log|\Sigma_y|$ is a concave, non-decreasing function in $\gamma$ and can be expressed as
\begin{eqnarray}
	\log|\Sigma_y| = \min_{z \geq 0} z^\top \gamma - h^*(z)
\end{eqnarray}
where $h^*(z)$ is the concave conjugate of $\log|\Sigma_y|$ and given by
\begin{eqnarray}
	h^*(z) = \min_{\gamma \geq 0 }z^\top\gamma - \log|\Sigma_y|.
\end{eqnarray}
Based on that, we rewrite \eqref{eq:g_fcn} as
\begin{eqnarray}
\label{eq:g_fcn_gamma_z}
	g(\theta) = \min_{\gamma, z \geq 0} \theta^\top \Gamma^{-1} \theta + z^\top \gamma - h^*(z).
\end{eqnarray}
Fixing $\theta$ and $z$, then optimizing over $\gamma$, yields
\begin{eqnarray}
	\gamma^{opt}_i = z_i^{-1/2}|\theta_i|,~~i=1,\ldots,m.
\end{eqnarray}
Substituting $\gamma^{opt}$ back to \eqref{eq:g_fcn_gamma_z}, we get
\begin{eqnarray}
	g(\theta) = \min_{z \geq 0}\sum_{i=1}^{m}2z_i^{-1/2}|\theta_i| - h^*(z).
\end{eqnarray}
Now, we only need to calculate the optimal $z$. For a fixed $\gamma$, we get 
\begin{eqnarray}
	z^{opt} &=& \nabla_\gamma \log |\Sigma_y| \\
	        &=& \text{diag}[\Phi^\top\Sigma_y^{-1}\Phi] = \text{diag}[\Phi^\top(\lambda I + \Phi\Gamma\Phi^\top)\Phi]\nonumber.
\end{eqnarray}
Then, we can substitute $\Gamma=\text{diag}[\gamma^{opt}]$ to get a value for $z^{opt}$. 

\bibliographystyle{ieeetran}
\bibliography{react_refs}

\begin{thebibliography}{10}
\providecommand{\url}[1]{#1}
\csname url@samestyle\endcsname
\providecommand{\newblock}{\relax}
\providecommand{\bibinfo}[2]{#2}
\providecommand{\BIBentrySTDinterwordspacing}{\spaceskip=0pt\relax}
\providecommand{\BIBentryALTinterwordstretchfactor}{4}
\providecommand{\BIBentryALTinterwordspacing}{\spaceskip=\fontdimen2\font plus
\BIBentryALTinterwordstretchfactor\fontdimen3\font minus
  \fontdimen4\font\relax}
\providecommand{\BIBforeignlanguage}[2]{{%
\expandafter\ifx\csname l@#1\endcsname\relax
\typeout{** WARNING: IEEEtran.bst: No hyphenation pattern has been}%
\typeout{** loaded for the language `#1'. Using the pattern for}%
\typeout{** the default language instead.}%
\else
\language=\csname l@#1\endcsname
\fi
#2}}
\providecommand{\BIBdecl}{\relax}
\BIBdecl

\bibitem{ljung_sysid}
L.~Ljung, \emph{System Identification: Theory for the User}.\hskip 1em plus
  0.5em minus 0.4em\relax Upper Saddle River, NJ, USA: Prentice Hall PTR, 1999.

\bibitem{Johnston2011conj}
M.~D. Johnston and D.~Siegel, ``Linear conjugacy of chemical reaction
  networks,'' \emph{Journal of Mathematical Chemistry}, vol.~49, pp.
  1263--1282, 2011.

\bibitem{acs2016}
B.~Ács, G.~Szederkényi, Z.~Tuza, and Z.~A. Tuza, ``Computing all possible
  graph structures describing linearly conjugate realizations of kinetic
  systems,'' \emph{Computer Physics Communications}, vol. 204, pp. 11--20,
  2016.

\bibitem{acs2018}
B.~Ács, G.~Szlobodnyik, and G.~Szederkényi, ``A computational approach to the
  structural analysis of uncertain kinetic systems,'' \emph{Computer Physics
  Communications}, vol. 228, pp. 83--95, 2018.

\bibitem{Feinberg:79}
M.~Feinberg, \emph{Lectures on chemical reaction networks}.\hskip 1em plus
  0.5em minus 0.4em\relax Notes of lectures given at the Mathematics Research
  Center, University of Wisconsin, 1979.

\bibitem{Feinberg1987}
------, ``Chemical reaction network structure and the stability of complex
  isothermal reactors - {I}. {T}he deficiency zero and deficiency one
  theorems,'' \emph{Chemical Engineering Science}, vol. 42 (10), pp.
  2229--2268, 1987.

\bibitem{Craciun2015}
G.~Craciun, ``Toric differential inclusions and a proof of the global attractor
  conjecture,'' January 2015, arXiv:1501.02860 [math.DS].

\bibitem{Shinar2010}
G.~Shinar and M.~Feinberg, ``Structural sources of robustness in biochemical
  reaction networks,'' \emph{Science}, vol. 327, pp. 1389--1391, 2010.

\bibitem{Craciun2009}
G.~Craciun, A.~Dickenstein, A.~Shiu, and B.~Sturmfels, ``Toric dynamical
  systems,'' \emph{Journal of Symbolic Computation}, vol.~44, pp. 1551--1565,
  2009.

\bibitem{perez2012chemical}
M.~P{\'e}rez~Mill{\'a}n, A.~Dickenstein, A.~Shiu, and C.~Conradi, ``Chemical
  reaction systems with toric steady states,'' \emph{Bulletin of mathematical
  biology}, vol.~74, no.~5, pp. 1027--1065, 2012.

\bibitem{van2013mathematical}
A.~van~der Schaft, S.~Rao, and B.~Jayawardhana, ``On the mathematical structure
  of balanced chemical reaction networks governed by mass action kinetics,''
  \emph{SIAM Journal on Applied Mathematics}, vol.~73, no.~2, pp. 953--973,
  2013.

\bibitem{Haddad2010}
W.~M. Haddad, V.~Chellaboina, and Q.~Hui, \emph{Nonnegative and Compartmental
  Dynamical Systems}.\hskip 1em plus 0.5em minus 0.4em\relax Princeton
  University Press, 2010.

\bibitem{GaborA2016}
A.~G{\'{a}}bor, K.~M. Hangos, and G.~Szederk{\'{e}}nyi, ``Linear conjugacy in
  biochemical reaction networks with rational reaction rates,'' \emph{Journal
  of Mathematical Chemistry}, vol.~54, no.~8, pp. 1658--1676, may 2016.

\bibitem{Erdi1989}
P.~Érdi and J.~Tóth, \emph{Mathematical Models of Chemical Reactions. Theory
  and Applications of Deterministic and Stochastic Models}.\hskip 1em plus
  0.5em minus 0.4em\relax Manchester, Princeton: Manchester University Press,
  Princeton University Press, 1989.

\bibitem{Hars1981}
V.~Hárs and J.~Tóth, ``On the inverse problem of reaction kinetics,'' in
  \emph{Qualitative Theory of Differential Equations}, ser. Coll. Math. Soc. J.
  Bolyai, M.~Farkas and L.~Hatvani, Eds.\hskip 1em plus 0.5em minus 0.4em\relax
  North-Holland, Amsterdam, 1981, vol.~30, pp. 363--379.

\bibitem{Craciun2008}
G.~Craciun and C.~Pantea, ``Identifiability of chemical reaction networks,''
  \emph{Journal of Mathematical Chemistry}, vol.~44, pp. 244--259, 2008.

\bibitem{Johnston2012a}
M.~D. Johnston, D.~Siegel, and G.~Szederkényi, ``Dynamical equivalence and
  linear conjugacy of chemical reaction networks: new results and methods,''
  \emph{MATCH Commun. Math. Comput. Chem.}, vol.~68, pp. 443--468, 2012.

\bibitem{TuzaFosbe2016}
Z.~A. Tuza, B.~{\'{A}}cs, G.~Szederk{\'{e}}nyi, and F.~Allgöwer, ``Efficient
  computation of all distinct realization structures of kinetic systems,''
  \emph{{IFAC}-{PapersOnLine}}, vol.~49, no.~26, pp. 194--200, 2016.

\bibitem{Szederkenyi2011b}
\BIBentryALTinterwordspacing
G.~Szederkényi, K.~M. Hangos, and Z.~Tuza, ``Finding weakly reversible
  realizations of chemical reaction networks using optimization,'' \emph{MATCH
  Commun. Math. Comput. Chem.}, vol.~67, pp. 193--212, 2012. [Online].
  Available: \url{http://arxiv.org/abs/1103.4741}
\BIBentrySTDinterwordspacing

\bibitem{Szederkenyi2011a}
G.~Szederkényi and K.~M. Hangos, ``Finding complex balanced and detailed
  balanced realizations of chemical reaction networks,'' \emph{Journal of
  Mathematical Chemistry}, vol.~49, pp. 1163--1179, 2011.

\bibitem{Liptak2015}
G.~Lipták, G.~Szederkényi, and K.~M. Hangos, ``Computing zero deficiency
  realizations of kinetic systems,'' \emph{Systems \& Control Letters},
  vol.~81, pp. 24--30, 2015.

\bibitem{Johnston2016}
M.~D. Johnston, ``A linear programming approach to dynamical equivalence,
  linear conjugacy, and the deficiency one theorem,'' \emph{Journal of
  Mathematical Chemistry}, vol.~54, no.~8, pp. 1612--1631, may 2016.

\bibitem{Szederkenyi2011}
G.~Szederkényi, K.~M. Hangos, and T.~Péni, ``Maximal and minimal realizations
  of reaction kinetic systems: computation and properties,'' \emph{MATCH
  Commun. Math. Comput. Chem.}, vol.~65, pp. 309--332, 2011.

\bibitem{Acs2015}
B.~Ács, G.~Szederkényi, Z.~A. Tuza, and Z.~Tuza, ``Computing linearly
  conjugate weakly reversible kinetic structures using optimization and graph
  theory,'' \emph{MATCH Commun. Math. Comput. Chem.}, vol.~74, pp. 481--504,
  2015.

\bibitem{liptak2016kinetic}
G.~Lipt{\'a}k, G.~Szederk{\'e}nyi, and K.~M. Hangos, ``Kinetic feedback design
  for polynomial systems,'' \emph{Journal of Process Control}, vol.~41, pp.
  56--66, 2016.

\bibitem{August2009}
E.~August and A.~Papachristodoulou, ``Efficient, sparse biological network
  determination,'' \emph{{BMC} Systems Biology}, vol.~3, p.~25, 2009.

\bibitem{wei2017phd}
W.~Pan, ``Bayesian learning for nonlinear system identification,'' Ph.D.
  dissertation, Imperial College London, 2017.

\bibitem{debrabanter2013}
\BIBentryALTinterwordspacing
K.~De~Brabanter, J.~De~Brabanter, B.~De~Moor, and I.~Gijbels, ``Derivative
  estimation with local polynomial fitting,'' \emph{J. Mach. Learn. Res.},
  vol.~14, no.~1, pp. 281--301, Jan. 2013. [Online]. Available:
  \url{http://dl.acm.org/citation.cfm?id=2567709.2502590}
\BIBentrySTDinterwordspacing

\bibitem{Schmidt2009}
M.~Schmidt and H.~Lipson, ``Distilling free-form natural laws from experimental
  data,'' \emph{Science}, vol. 324, no. 5923, pp. 81--85, apr 2009.

\bibitem{Brunton2016}
S.~L. Brunton, J.~L. Proctor, and J.~N. Kutz, ``Discovering governing equations
  from data by sparse identification of nonlinear dynamical systems,''
  \emph{Proceedings of the National Academy of Sciences}, vol. 113, no.~15, pp.
  3932--3937, mar 2016.

\bibitem{Tipping2001}
M.~E. Tipping, ``Sparse bayesian learning and the relevance vector machine,''
  \emph{Journal of machine learning research}, vol.~1, no. Jun, pp. 211--244,
  2001.

\bibitem{Candes2008}
E.~J. Candes, M.~B. Wakin, and S.~P. Boyd, ``Enhancing sparsity by reweighted
  l1 minimization,'' \emph{Journal of Fourier analysis and applications},
  vol.~14, no. 5-6, pp. 877--905, 2008.

\bibitem{Wipf2010}
D.~Wipf and S.~Nagarajan, ``Iterative reweighted l1 and l2 methods for finding
  sparse solutions,'' \emph{IEEE Journal of Selected Topics in Signal
  Processing}, vol.~4, no.~2, pp. 317--329, 2010.

\bibitem{Wipf2011}
D.~P. Wipf, B.~D. Rao, and S.~Nagarajan, ``Latent variable bayesian models for
  promoting sparsity,'' \emph{IEEE Transactions on Information Theory},
  vol.~57, no.~9, pp. 6236--6255, 2011.

\bibitem{Wipf2008}
D.~P. Wipf and S.~S. Nagarajan, ``A new view of automatic relevance
  determination,'' in \emph{Advances in neural information processing systems},
  2008, pp. 1625--1632.

\bibitem{Wipf2009}
D.~Wipf and S.~Nagarajan, ``A unified bayesian framework for meg/eeg source
  imaging,'' \emph{NeuroImage}, vol.~44, no.~3, pp. 947--966, 2009.

\bibitem{palmer2006variational}
J.~Palmer, K.~Kreutz-Delgado, B.~D. Rao, and D.~P. Wipf, ``Variational em
  algorithms for non-gaussian latent variable models,'' in \emph{Advances in
  neural information processing systems}, 2006, pp. 1059--1066.

\bibitem{wipf2007performance}
D.~Wipf, J.~Palmer, B.~Rao, and K.~Kreutz-Delgado, ``Performance analysis of
  latent variable models with sparse priors,'' in \emph{Proceedings of ICASSP
  2007}, 2007.

\bibitem{Boyd2004}
S.~Boyd and L.~Vandenberghe, \emph{Convex Optimization}.\hskip 1em plus 0.5em
  minus 0.4em\relax Cambridge University Press, 2004.

\end{thebibliography}

\end{document}